\documentclass[twocolumn,english,aps,prl,showpacs]{revtex4}
\usepackage[T1]{fontenc}
\usepackage{amsmath,graphicx,amssymb,epsfig,babel,dsfont,color}

\newcommand{\rt}{{\rm t}}
\newcommand{\kb}{k_{\rm B}}

\begin{document}

\title{Radiative bistability and thermal memory}

\author{Viacheslav Kubytskyi$^{1}$, Svend-Age Biehs$^{2}$ and Philippe Ben-Abdallah$^{1,*}$}

\affiliation{$^1$Laboratoire Charles Fabry,UMR 8501, Institut d'Optique, CNRS, Universit\'{e} Paris-Sud 11,
2, Avenue Augustin Fresnel, 91127 Palaiseau Cedex, France.}
\email{pba@institutoptique.fr}
\affiliation{$^2$Institut f\"{u}r Physik, Carl von Ossietzky Universit\"{a}t, D-26111 Oldenburg, Germany.}

\date{\today}


\begin{abstract}
We predict the existence of a thermal bistability in many-body systems out of thermal 
equilibrium which exchange heat by thermal radiation using insulator-metal transition (IMT) materials. 
We propose a writing-reading procedure and demonstrate the possibility to exploit the thermal 
bistability to make a volatile thermal memory. We show that this thermal memory can be used to 
store heat and thermal information (via an encoding temperature) for arbitrary long times. 
The radiative thermal bistability could find broad applications in the domains of thermal 
management, information processing and energy storage.

\end{abstract}

\maketitle

The control of electric currents with diodes and transitors is undoubtedly a cornerstone in 
modern electronics which has revolutionized our daily life. Astonishingly, similar devices
which allow for controlling the heat flow are not as widespread as their electronic counterparts. 
In 2006 Baowen Li et al.~\cite{Casati1} have proposed a thermal analog of a field-effect transistor 
by replacing both the electric potentials and the electric currents in the electronic circuits by 
thermostats and heat fluxes carried by phonons through solid segments. Few years later, several 
prototypes of phononic thermal logic gates~\cite{BaowenLi2} as well as thermal 
memories~\cite{BaowenLi3,BaowenLiEtAl2012} have been developed from these basic building blocks 
allowing for processing information with heat currents rather than with electric currents. 
However, this phonon-based technology is intrinsically limited by the speed of the heat carriers, 
the acoustic phonons, and by the presence of Kapitza resistances between the basic solid elements.  
Moreover, because of radiative losses and thermal fluctuations, the temporal stability of these 
systems is limited and a frequent refreshing is needed. To overcome these problems optical contactless analogs
of some basic devices such as the radiative thermal diode~\cite{OteyEtAl2010,PBA_APL} and 
the radiative transistor~\cite{PBA_PRL2014} have been recently introduced. However some 
functionalities are to date missing to allow for an all-photonic treatment of heat fluxes. 

In this Letter, we make a step forward in this direction by introducing the concept of a radiative 
thermal memory which is able to store information for arbitrary long time using thermal photons. 
To do so, we first demonstrate the existence of bistable thermal behavior in simple systems consisting 
of two membranes which are out of thermal equilibrium and which are further sandwiched between two thermal baths
at different temperatures. Existence of multiple equilibrium temperature requires~\cite{BaowenLi3} the 
presence of negative differential thermal resitances (NDTR). However, as shown by Fan et al.~\cite{Fan_APL} this is not 
at all a sufficient condition. As we will show, the thermal bistability mechanism can only exist 
in many-body systems \cite{PBA_PRL2011,Messina,Messina2}. Finally, as direct consequence, we show that the 
bistability can be used to store one bit of thermal information similar. While in a conventional electronic memory, 
the states "$0$" and "$1$" are defined by two different applied voltages for which the electric current 
inside the circuit is zero, its thermal counterpart is defined with distinct equilibrium temperatures 
wich lead to vanishing heat fluxes between the different parts of the system.

\begin{figure}[Hhbt]
\includegraphics[scale=0.35]{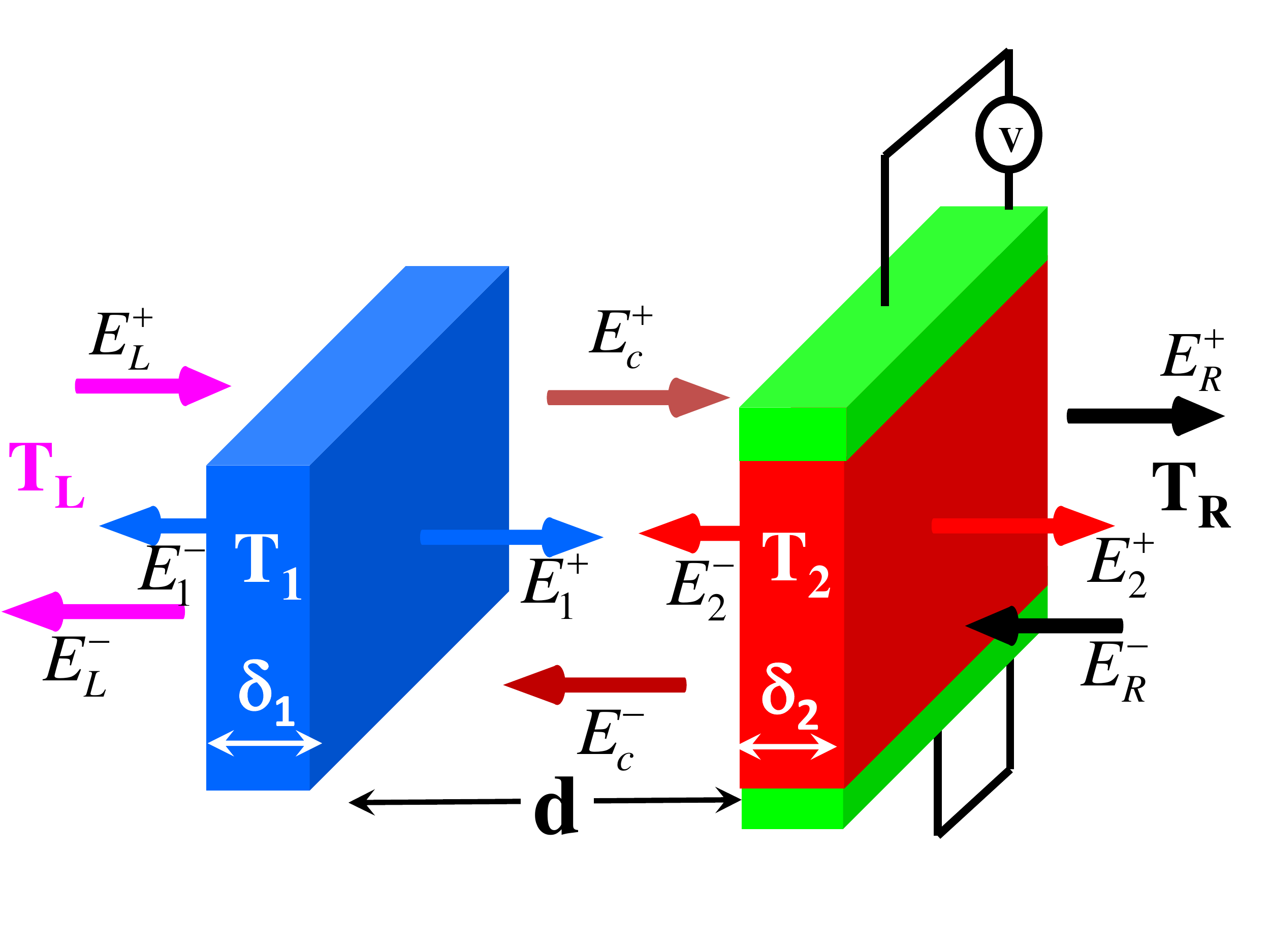}
\caption{Sketch of a radiative thermal memory. A membrane made of an IMT material is placed at a distance 
$d$ from a dielectric layer. The system is surrounded by two thermal baths at different temperatures $T_{\rm L}$ and $T_{\rm R}$. 
The temperature $T_2$ can be increased or reduced either by Joule heating by applying a voltage difference through a couple of 
electrodes or by using Peltier elements.} 
\end{figure} 

Let us consider a system as depicted in Fig.~1 composed by two parallel homogeneous membranes 
of finite thicknesses $\delta_{1}$ and $\delta_{2}$  and separated by a distance $d$. The left (right) membrane
is in contact with a thermal bath having temperature $T_{\rm L}$ ($T_{\rm R}$), where $T_{\rm L} \neq T_{\rm R}$. 
The membranes themselves interact on the one hand through the intracavity fields and on the other with the thermal 
baths which can be thought to be produced by external media. In that sense, the system is driven by many-body interactions. 

\begin{figure}[Hhbt]
\includegraphics[scale=0.4, angle=-90]{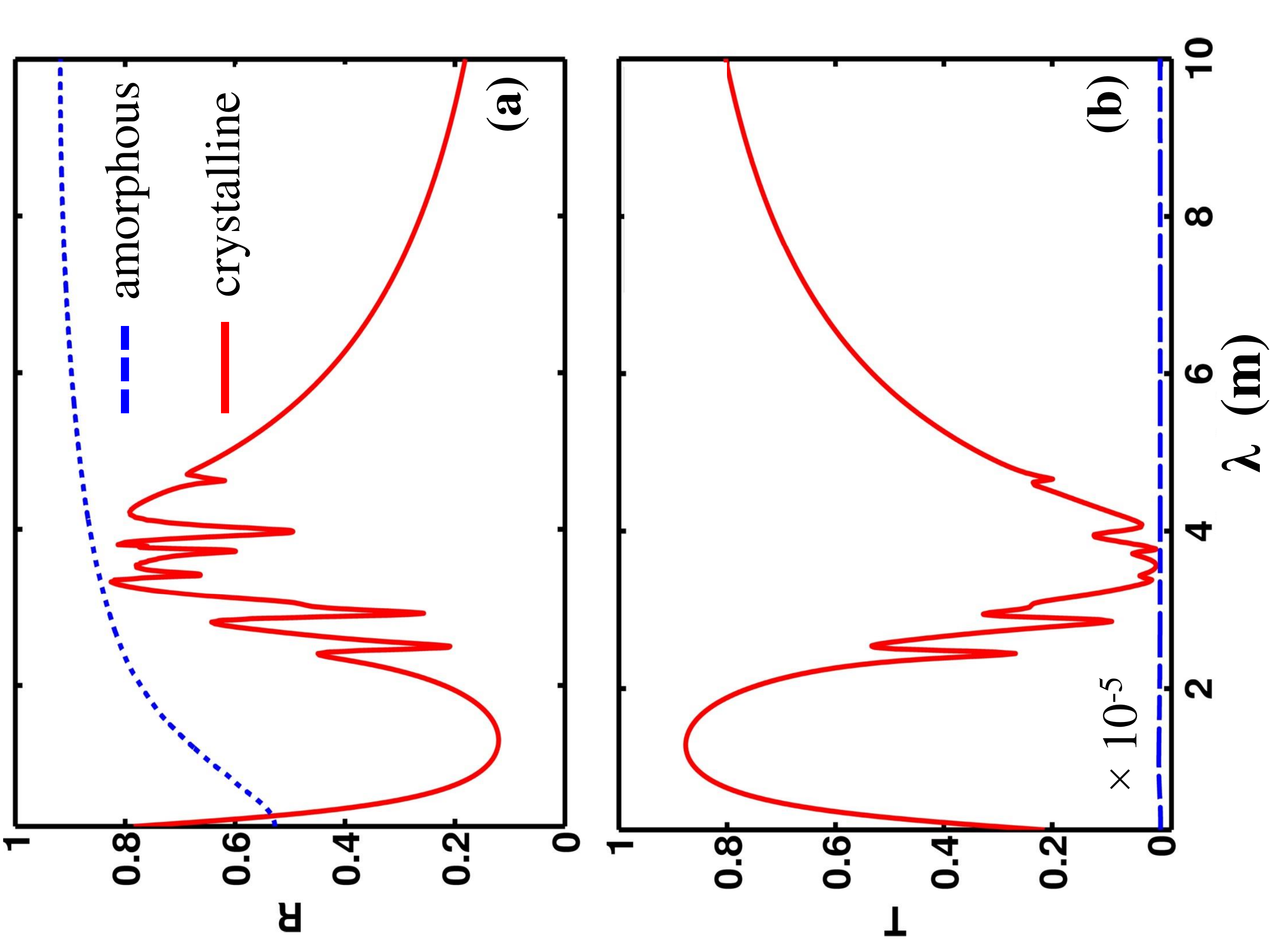}
\caption{Reflectance (a) and transmittance (b) at normal incidence of a $1\mu m$ thick VO$_2$ membrane in its amorphous (metallic) and crystalline (dielectric) phase.} 
\end{figure} 

The heat flux across any plane $z=\bar{z}$ parallel to the interacting surfaces is given by the normal component of Poynting vector
\begin{equation}
  \varphi(\bar{z})= \langle \mathbf{E}(\bar{z})\times\mathbf{H}(\bar{z})\rangle \cdot \boldsymbol{e}_z.
  \label{Eq:flux}
\end{equation}
Here $\boldsymbol{e}_z$ denotes the unit vector normal to the interfaces of the membranes; $\mathbf{E}$ and $\mathbf{H}$ are the 
local electric and magnetic fields which, according to fluctuational electrodynamics developed by Rytov~\cite{Rytov,Polder1973}, 
are either generated by randomly fluctuating source currents in both membranes or given by the fluctuational fields of 
the thermal baths. Here $\langle \circ \rangle$ represents the classical statistical averaging over all field realizations. 
For the sake of simplicity we assume that the separation distance $d$ is large enough compared to the thermal 
wavelengths [i.e. $d\gg\max(\lambda_{T_i}=c\hbar/(\kb T_{i})$, $i = 1,2,{\rm L}, {\rm R}$ ] so that near-field heat exchanges 
can be neglected. Then, according to the theory of fluctuational electrodynamics the heat flux can be written in terms 
of the field correlators $\mathfrak{C}_j^{\phi,\phi'}(\omega,\boldsymbol{\kappa})= \frac{1}{2} \langle[E_j^\phi(\omega,\boldsymbol{\kappa})E_j^{\phi'\dagger}(\omega,\boldsymbol{\kappa})+E_j^{\phi'\dagger}(\omega,\boldsymbol{\kappa})E_j^\phi(\omega,\boldsymbol{\kappa})]\rangle$ of local field amplitudes in polarization $j$~\cite{Messina3}
\begin{equation}
\begin{split}
     \varphi(\bar{z}) = 2\epsilon_0c^2 \!\!\sum \limits_{\underset{\phi=\left\{ +,-\right\}}{j=s,p}}\int_{0}^{\infty}\!\!\frac{d\omega}{2\pi}\int\!\!\! \frac{{\rm d}^2 \boldsymbol{\kappa}}{(2 \pi)^2}\frac{\phi k_z}{\omega} \mathfrak{C}_j^{\phi,\phi}(\omega,\boldsymbol\kappa),
\label{Eq:base}
\end{split}
\end{equation}
where $ \boldsymbol{\kappa}=(k_x,k_y)$ and $k_z=\sqrt{\frac{\omega^2}{c^2}-\boldsymbol{\kappa}^2}$ denote the parallel and normal 
components of the wavector.  The local field and therefore the correlator $\mathfrak{C}_j^{\phi,\phi}$ can be expressed 
(see \cite{SupplMat} for details) after scattering in terms of fields $E^{\pm}_i$ ($i = 1,2,{\rm L},{\rm R}$) emitted 
by each body and by both thermal reservoirs as illustrated in Fig~1. Using expression (\ref{Eq:base}) we can calculate the
net flux $\Phi_1=\varphi(0)-\varphi(-\delta_1)$ [$\Phi_2=\varphi(d+\delta_2)-\varphi(d)$] received by the first (second) membrane.

Now, let us assume that one of both membranes, medium 2, say, is made of vanadium dioxide (VO$_2$) an IMT material which 
undergoes a first-order transition (Mott transition~\cite{Mott}) from a high-temperature metallic phase to a low-temperature 
insulating phase~\cite{Baker} at a critical temperature $T_{\rm c}$ close to room-temperature ($T_{\rm c}=340\,{\rm K}$). As shown in Fig.~2, during the phase transition
of VO$_2$ which can be initiated by a small change of its temperature around $T_c$ the optical 
properties of VO$_2$ change drastically. Different works have already shown~\cite{PBA_APL,van Zwol2} that the exchanged heat-flux 
is drastically affected by this transition. As far as medium 1 is concerned, we use a SiO$_2$ membrane~\cite{Palik} which is 
partially transparent in the infrared range. Hence, depending on the crystaline phase of IMT, the membranes are either under the 
influence of both thermal reservoirs or mainly of one of them, since the VO$_2$ layer in its metallic phase is playing the role 
of a mirror in the infrared as can be seen in Fig.~2.

The time evolution of the temperatures $T_1$ and $T_2$ of the membranes are solution of the following nonlinear 
coupled system of differential equations
\begin{equation}
  \partial_t \mathbf{T} = \boldsymbol{\Phi} + \mathbf{Q}
  \label{Eq:diff}
\end{equation}
where we have introduced the vectors $\mathbf{T} \equiv \bigl(T_1(t),T_2(t)\bigr)^\rt$, $\boldsymbol{\Phi} \equiv \bigl(\Phi_1(T_1,T_2)/I_1, \Phi_2(T_1,T_2)/I_2\bigr)^\rt$,
and $\mathbf{Q} \equiv (Q_1\delta_1/I_1,Q_2\delta_2/I_2)^\rt$. Here  $Q_i$ ($i = 1,2$) is the power per unit volume which can be added to or extracted from both 
membranes by applying a voltage difference through a couple of electrodes as illustrated in Fig.~1 or by using Peltier elements.
Furthermore, we have introduced the thermal inertia of both membranes as $I_i \equiv C_i \rho_i \delta_i$, where $C_i$ and $\rho_i$ are the 
heat capacity and the mass density of each material. By writing down this set of equations we have neglected any temperature variation
inside the membranes which is a very good approximation given that the conductivity inside the membranes is much larger than between 
the membranes. When assuming that no energy is directly added to or removed from the membranes, then $\mathbf{Q} = \mathbf{0}$. In this case, 
the steady-state solution is given by $\boldsymbol{\Phi} = \mathbf{0}$. Hence $\Phi_1$ and $\Phi_2$ vanish for the same couple of
temperatures $(T_1^{({\rm st-st})},T_2^{(\rm st-st)})$. Considering for an instant membrane 2 only, then the existence of two equilibrium 
temperatures where the net flux vanishes ($\Phi_2 = 0$) implies that $\Phi_2$ must have a maximum or a minimum between these two temperatures.
Hence, this requires a  negative differential resistive behavior for this membrane~\cite{Fan_APL} which was shown for IMT materials as
 VO$_2$~\cite{PBA_PRL2014}. For the whole system it is therefore a precondition to have at least one membrane which exihibits negative 
differential resistive behavior (i.e.\ VO$_2$) in order to have two couples  $(T_1^{({\rm st-st})},T_2^{(\rm st-st)})$ of steady-state temperatures.

In Figs. 3(a) and 3(b), we show the time evolutions of two SiO$_2$/VO$_2$ systems without external excitation (i.e. $\mathbf{Q} = \mathbf{0}$) 
when the thermal inertia $I_i$ of both membranes are comparable (i.e. $\delta_1\thickapprox\delta_2$) and very different (i.e. $\delta_1\gg\delta_2$). 
The trajectories (the thick pink and turquois lines) are obtained by solving Eq.~(\ref{Eq:diff}) using a Runge-Kutta method with adaptative time 
steps choosing different initial conditions. In these figures, the dashed blue (solid red) line represents the local equilibrium temperatures for 
the first (second) membrane that is the set of temperatures couples ($T_1,T_2$) which satisfy the condition $\Phi_1(T_1,T_2)=0$ [$\Phi_2(T_1,T_2)=0$]. 
The intersection of these two lines define the global steady-state temperatures of the system where $\boldsymbol{\Phi} = \mathbf{0}$. In Fig.~3 we 
observe three global steady-state temperature couples $(T_1^{(l)},T_1^{(l)})$ with $l = 1,2,3$ for both configurations. 

The stability of these steady-state temperatures can be deduced from an analysis of the eigenvalues of the Jacobian of the vector field $\boldsymbol{\Phi}$.
We find~\cite{SupplMat} that  $(T^{(1)}_1,T^{(1)}_2)$ and $(T^{(3)}_1,T^{(3)}_2)$ are fixed points, whereas $(T^{(2)}_1,T^{(2)}_2)$ is a
saddle point. It is worth noting that at the fixed points even in presence of thermal fluctuations the system remains in the steady state.
Furthermore, it is interesting to see how the system dynamic changes with respect to the thermal inertia $I$ of membranes. When $I_1\sim I_2$ we see in Fig.~3(a) 
that both membranes simultaneously cool down or heat up towards one of stable states. On the other hand, if one membrane has a strong inertia with
respect to the second one ($I_1\gg I_2$) we find two time scales for the relaxation towards the steady state, as illustrated in Fig.~3(b). First, the 
membrane with the smaller inertia reaches its local equilibrium state [defined by $\Phi_2=0$ in Fig.~3(b)] by bypassing the unstable state as 
shown in Fig.~3(b) and then the whole system relaxes toward a global stable state.

\begin{figure}[Hhbt]
\includegraphics[scale=0.4, angle=-90]{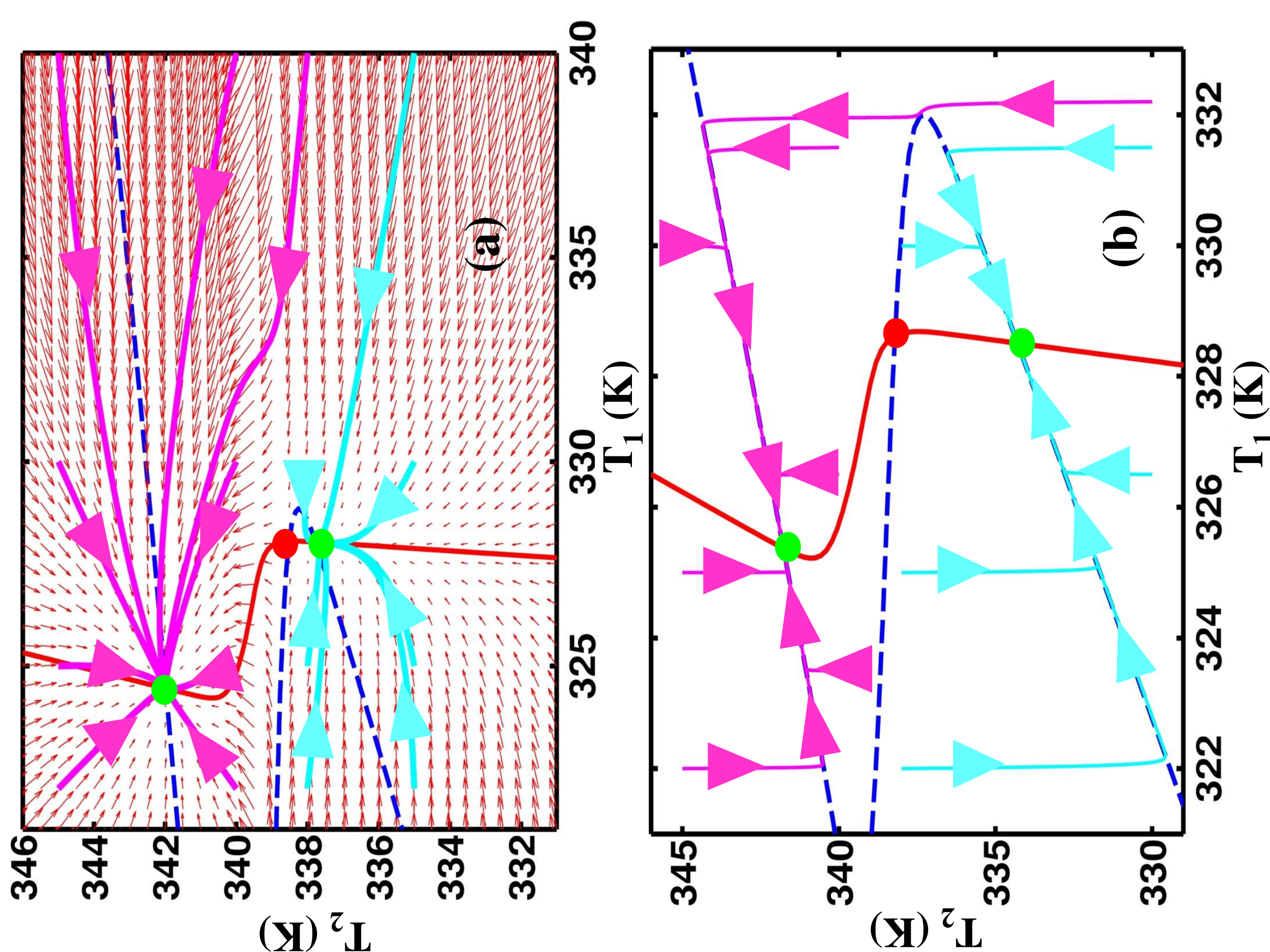}
\caption{ (a) Trajectories of temperatures (pink and turquois lines) for different initial conditions in the plane ($T_1,T_2$) in a two membrane SiO$_2$/VO$_2$ system with $\delta_1=\delta_2=1\,\mu{\rm m}$. 
The  blue dashed and red solid lines represent the local equilibrium conditions $\Phi_1 = 0$ and $\Phi_2 = 0$ of each membrane. The green (red) points denote the stable (unstable) 
global steady-state temperatures, $(T_1^{(1)},T_2^{(1)}) = (328.03\,{\rm K},337.77\,{\rm K})$, $(T_1^{(2)},T_2^{(2)}) = (328.06\,{\rm K},338.51\,{\rm K})$, and  $(T_1^{(3)},T_2^{(3)}) = (324.45\,{\rm K},341.97\,{\rm K})$.
The red arrows represents the vector field $\boldsymbol{\Phi}$. The temperature of thermal reservoirs are $T_{\rm L}=320\,{\rm K}$ and $T_{\rm R}=358\,{\rm K}$. 
(b) Temperature trajectories as in (a) but for a two-membrane SiO$_2$/VO$_2$ system with $\delta_1=1\,{\rm mm}$ and $\delta_2=1\,\mu{\rm m}$ choosing $T_{\rm L}=320\,{\rm K}$ and $T_{\rm R}=355{\rm K}$. 
In both configurations, the separation distance $d$ is much larger than the thermal wavelengths.} 
\end{figure} 

So far we have identified the stable thermal states $(T^{(1)}_1,T^{(1)}_2)$ and $(T^{(3)}_1,T^{(3)}_2)$ which can
be regarded as the two states "$0$" and  "$1$" of a bit. Now, we want to examine the transition between these two 
states. To switch from one thermal state to the other, we need to add or extract power from the system. In the following 
we describe this writing-reading procedure. To this end, we consider the SiO$_2$/VO$_2$ system made with membranes of equal thicknesses 
$\delta_1 = \delta_2=1\,\mu{\rm m}$ which is coupled to two reservoirs of temperatures $T_L=320 K$ and $T_R=358 K$. 
Let us define "0" as the thermal state at the temperature $T_2=\min(T^{(1)}_2,T^{(3)}_2)$. To make the transition towards the thermal state "1" 
the VO$_2$ membrane must be heated. 

\begin{figure}[Hhbt]
\includegraphics[scale=0.4, angle=-90]{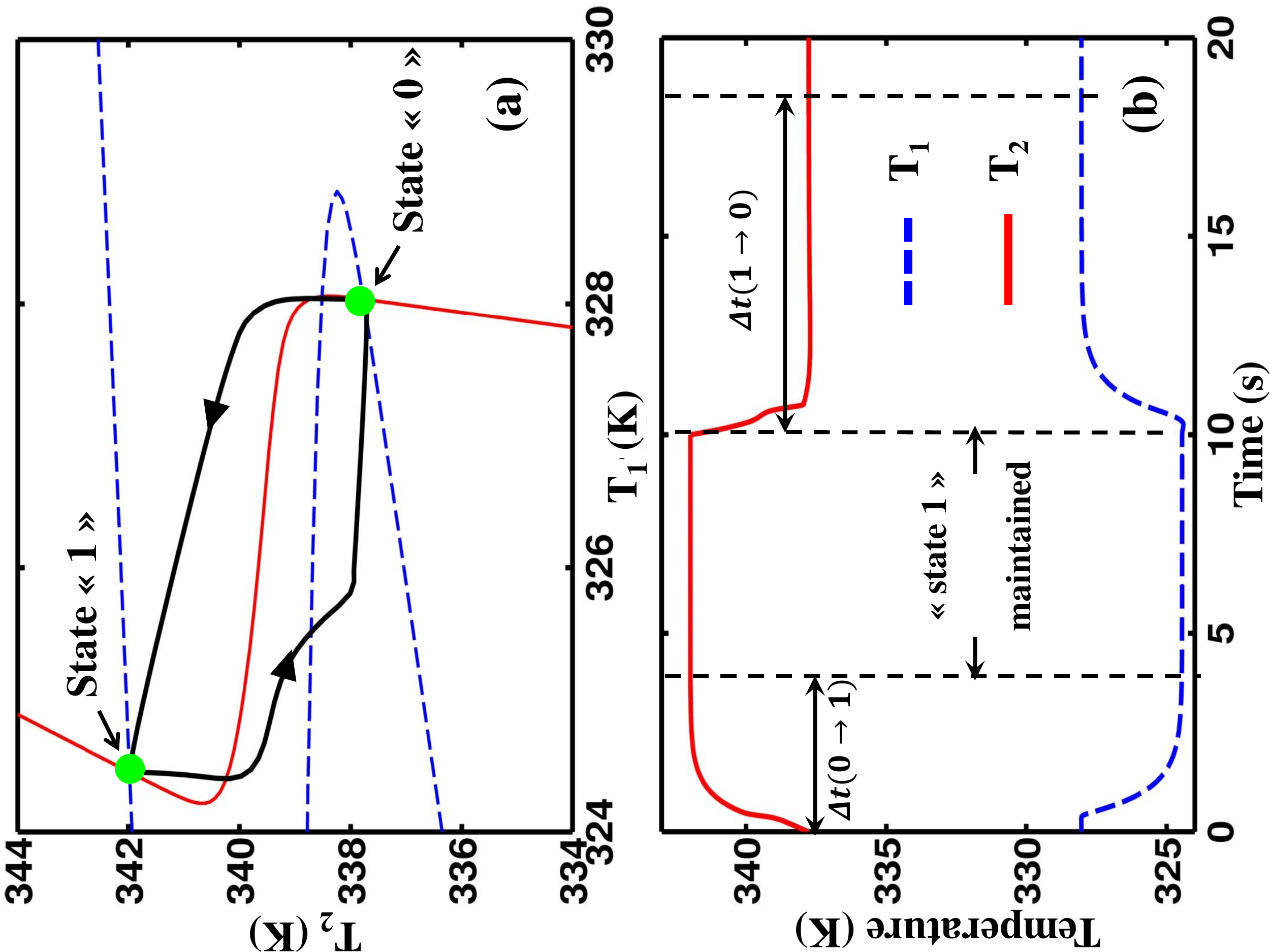}
\caption{(a) Hystereis of the VO$_2$ membrane temperature during a transition between the thermal states "0" and  "1" inside a two membrane SiO$_2$/VO$_2$ system 
with $\delta_1=\delta_2=1\,\mu{\rm m}$. The volumic powers supplied and extracted from the VO$_2$ layer during a time interval $\Delta t_1=0.4\,{\rm s}$ and $\Delta t_2=1.5\,{\rm s}$  
are $Q_2=10^{-2}\,{\rm W} {\rm mm}^{-3}$ and $Q_2=-2.5\times10^{-2}{\rm W}{\rm mm}^{-3}$, respectively. The writing time of state "1" ("0") from the state "0" ("1") is  
$\Delta t(0\rightarrow 1)=4\,{\rm s}$ ($\Delta t(1\rightarrow 0)=8\,{\rm s}$). (b) Time evolution $T_1(t)$ and $T_2(t)$ of SiO$_2$ and VO$_2$ membrane temperatures. 
The thermal states "0" and "1" can be maintained for arbitrary long time provided that the thermostats ($T_L=320\,{\rm K}$ and $T_R=358\,{\rm K}$) remain switched on.\label{Fig:Switch}} 
\end{figure}

Step 1 (transition from the state "0" to the state "1"): A volumic power $Q_2=10^{-2}\,{\rm W}{\rm mm}^{-3}$ 
is added to this membrane during a time interval $\Delta t_1\thickapprox 0.4\,{\rm s}$ to reach a region in the plane ($T_1,T_2$) [see Fig.~\ref{Fig:Switch}(a)]
where all trajectories converge naturally (i.e.\ for $Q_2=0$) after some time toward the state "1", the overall transistion time 
is $\Delta t(0\rightarrow 1)=4 s$ [Fig.~\ref{Fig:Switch}(b)].


Step 2 (maintaining the stored thermal information): Since the state "1" is a fixed point, the thermal data can be maintained for arbitrary long time provided that the thermal reservoirs are switched on. 
This corresponds basically to the concept of volatile memory in electronics.  

Step 3 (transition from the state "1" to the state "0"): Finally, a volumic power $Q_2=-2.5\times 10^{-2}\,{\rm W}{\rm mm}^{-3}$ is extracted from the VO$_2$ 
membrane during a time interval $\Delta t_2\thickapprox 1.5 s$ to reach a region [below $T_2=338\,{\rm K}$ in Fig.~\ref{Fig:Switch}(a)] of natural convergence to the state "0" . In this case the transition 
time becomes $\Delta t(1\rightarrow 0)=8\,{\rm s}$. Compared with its heating, the cooling of VO$_2$ does not follows the same trajectory [see Fig.~\ref{Fig:Switch}(a)] outlining 
the hysteresis of system which accompanies its bistable behavior. To read out the thermal state of system a classical electronic thermometer based on the thermo 
dependance of the electric resistivity of membranes can be used.

In conclusion, we have predicted the existence of bistable thermal behaviors in many-body systems in 
mutual radiative interaction and we have demonstrated the feasability for contactless thermal analogs 
of volatile electronic memories based on this effect which do not require refreshing. In this Letter 
our proofs of principle have been established in the far-field regime. Its generalization to near-field 
regime is straighforward. Since at short separation distances, thanks to photon tunneling, the heat flux exchanged between both 
membranes is not limited by the Stefan-Boltzmann law anymore, but can be increased by orders of magnitude, the speed of 
natural cooling/heating could also be drastically increased. We think that the 
radiative thermal memories pave the way for a contactless treatment of heat flows. They could find 
broad applications in the domains of thermal management, information processing and energy storage. 

%
%

\begin{acknowledgments}
The authors acknowledge financial support by the DAAD and Partenariat Hubert Curien Procope Program (project 55923991).
P.B.-A. acknowledges financial support by the Agence Nationale de la Recherche through the Source-TPV project ANR 2010 BLANC 0928 01 and thanks Prof. H. Benisty for fruitfull discussions.
\end{acknowledgments}

\end{document}